\documentstyle[amstex,preprint,aps]{revtex}
\tightenlines
\begin{document}
\title{Rigid rotor in phase space}
\author{S. Danko Bosanac}
\address{R. Boskovic Institute\\
10001 Zagreb, Croatia}
\maketitle

\begin{abstract}
Angular momentum is important concept in physics, and its phase space
properties are important in various applications. In this work phase space
analysis of the angular momentum is made from its classical definition, and
by imposing uncertainty principle its quantum properties are obtained. It is
shown that kinetic energy operator is derived, but it has different
interpretation of its parts than in the standard treatment. Rigid rotor is
discussed and it is shown what is its phase space representation. True rigid
rotor is defined and also its phase space properties are discussed.
\end{abstract}

\section{Introduction}

Angular momentum is a very important concept in dynamics of particles and
within quantum mechanics its properties are very well understood \cite{zare}%
. One example where the theory has direct application is the rigid rotor
model, which is basic for understanding rotational spectroscopy and
collisions of molecules. These processes are described by quantum dynamics,
but there are circumstances when classical dynamics is used as alternative.
For example, rotational cross sections for two colliding molecules in
principle can be calculated from quantum mechanics, but often it is a
challenging task. Classical mechanics,on the other hand, is relatively
simple to use, but there several problems in its implementation, say to
calculate atom-molecule collision cross sections. The basic one is how the
initial conditions are selected and the final results analyzed. For example,
if one says that a molecule is in the rotational state with the quantum
numbers $l=3$ and $m=2$ then the question is what to choose for the initial
orientation and angular velocity which adequately represents it? Analogous
problem had been analyzed for collisions involving only vibrational energy
exchange in atom-molecule collisions, where it was shown how to select \
initial conditions that adequately represent a particular vibrational state
of the molecule. A recipe was suggested from which transition probabilities
were successfully calculated \cite{bos2} from classical mechanics, and by
that it is meant that initial conditions were selected from a prescribed
phase space density but dynamics is calculated from classical equations of
motion.

The problem is therefore how does one chooses a set of initial positions and
velocities for a particle so that on average its angular momentum has given $%
l$ and $m$ values? In other words, what is needed is a function $f_{l,m}(%
\vec{r},\vec{p})$ with a property that the following averages are obtained 
\begin{eqnarray}
&<&\left( \vec{r}\times \vec{p}\right) ^{2}>=\int d^{3}r\;d^{3}p\;\left( 
\vec{r}\times \vec{p}\right) ^{2}f_{l,m}(\vec{r},\vec{p})=\hbar ^{2}l(l+1)
\label{claver} \\
&<&\vec{r}\times \vec{p}>=\int d^{3}r\;d^{3}p\;\vec{r}\times \vec{p}%
\;f_{l,m}(\vec{r},\vec{p})=\hbar m\;\hat{z}  \nonumber
\end{eqnarray}
In classical mechanics the function $f_{l,m}(\vec{r},\vec{p})$ is
interpreted as a statistical weight from which the position and momentum are
selected. It should be noted that it is not necessary that $f_{l,m}(\vec{r},%
\vec{p})$ is positive function, as long as those averages are obtained. In
practice this means that initial conditions are randomly selected from $%
\left| f_{l,m}(\vec{r},\vec{p})\right| $ and if $N$ sets are chosen then the
averages are approximately 
\[
<\left( \vec{r}\times \vec{p}\right) ^{2}>\approx \frac{\sum_{j=1}^{N}\left( 
\vec{r}_{j}\times \vec{p}_{j}\right) ^{2}\;sign\left[ f_{l,m}(\vec{r}_{j},%
\vec{p}_{j})\right] }{\sum_{j=1}^{N}sign\left[ f_{l,m}(\vec{r}_{j},\vec{p}%
_{j})\right] } 
\]
and similarly for the angular momentum. The \ function $sign\left[ f_{l,m}(%
\vec{r}_{j},\vec{p}_{j})\right] $ is the sign of $f_{l,m}(\vec{r}_{j},\vec{p}%
_{j})$ for the set of initial conditions $(\vec{r}_{j},\vec{p}_{j})$.
However, much more stringent condition is that the function $f_{l,m}(\vec{r},%
\vec{p})$ \ is stationary. The meaning of this can demonstrated on simple
example. If the particle of mass $M$ is free and the set of its initial
conditions is $(\vec{r}_{j},\vec{p}_{j})$ then its position after time $t$
is $\vec{r}=\vec{r}_{j}+\frac{\vec{p}_{j}}{M}t$ and velocity $\vec{v}=\frac{%
\vec{p}_{j}}{M}$. From these values one would be able to calculate a new
distribution function $g(\vec{r},\vec{p})$, by requiring to have properties
as previously described for the function $f_{l,m}(\vec{r},\vec{p})$.
Stationarity requires that the two functions are identical for any time $t$.
This restriction is very important, for obvious reasons, and it is
sufficient to dismiss a large number of ad-hoc distributions.

The most obvious starting point would be the wave function for a particle in
a particular angular momentum state, which in the coordinate space is $\psi
_{l,m}(\vec{r})=\chi _{n,l}(r)Y_{l,m}(\theta ,\phi )$, where $Y_{l,m}(\theta
,\phi )$ is spherical harmonic. Likewise in momentum space the wave function
is $\varphi _{l,m}(\vec{p})=\omega _{n,l}(p)Y_{l,m}(\theta _{p},\phi _{p})$,
where $\theta _{p}$ and $\phi _{p}$ are angles of the momentum. One could
then form the function $f_{l,m}(\vec{r},\vec{p})=\left| \chi
_{n,l}(r)Y_{l,m}(\theta ,\phi )\right| ^{2}\left| \omega
_{n,l}(p)Y_{l,m}(\theta _{p},\phi _{p})\right| ^{2}$ from which initial
position vector and momentum could be selected, except that neither the
averages (\ref{claver}) are obtained nor the condition of stationarity for
the function $f_{l,m}(\vec{r},\vec{p})$ is satisfied, which is easily
proven. Therefore this simple idea fails, and this is because one basic
difference between classical and quantum dynamics was not taken into
account: classical dynamics is defined in the phase space, while quantum
dynamics is either in the coordinate or momentum space. Simple product that
was suggested for the distribution $f_{l,m}(\vec{r},\vec{p})$ only reflects
this: it says that initial conditions in the phase space are determined as
the product of the distribution from the coordinate {\it and }momentum
space, as being independent. What is needed is formulation of quantum
mechanics in the phase space, and then one would possibly be able to satisfy
those two conditions. The earliest attempt in this direction was done by
Wigner\cite{wigner} who looked for a function $\rho (\vec{r},\vec{p},t\,)$
with the property 
\begin{equation}
P(\vec{r},t\,)=\left| \psi (\vec{r},t\,)\right| ^{2}=\int d^{3}p\;\rho (\vec{%
r},\vec{p},t\,)\;\;\;;\;\;\;\;Q(\vec{p},t\,)=\left| \varphi (\vec{p}%
,t\,)\right| ^{2}=\int d^{3}r\;\rho (\vec{r},\vec{p},t\,)  \label{pq}
\end{equation}
and obtained, what is known as, the Wigner function 
\begin{equation}
\rho (\vec{r},\vec{p},t\,)\;=\;\frac{1}{\pi ^{3}\hbar ^{3}}\int d^{3}q\,e^{2i%
\vec{p}\cdot \vec{q}/\hbar }\,\psi ^{\ast }(\vec{r}+\vec{q}\,,t)\psi (\vec{r}%
-\vec{q},t\,)  \label{wign}
\end{equation}
which is regarded as extension of quantum mechanics into the phase space. By
defining it in this way inevitably results in non uniqueness of the
extension \cite{lee}, because there are a large class of phase space
functions that are defined by the requirements (\ref{pq}). Another attempt
to formulate quantum mechanics in the phase space is due to Moyal\cite
{moyal,gadella} whose main objective was to give it a sound statistical
foundation. The main starting point is postulating observables as operators,
and postulating that to each set of commutating observables there is a set
of non commuting. Together they form a complete set of observables. For
these observable (operators) one forms a function from which the
characteristic function of statistical theory for a given quantum state $%
\psi $ is obtained as a matrix element. By the Fourier transform Moyal
obtains distribution which specifies to the Wigner function if the
observables are coordinates and momenta. Common to both approaches is that
the principles of quantum mechanics are assumed, although Moyal analysis is
more general. Can the Wigner function be used for the distribution $f_{l,m}(%
\vec{r},\vec{p})$? It is not clear that the averages (\ref{claver}) are
obtained, in fact as it will be shown later they are only partly obtained.
Furthermore, it is also not clear that the stationarity condition is
satisfied, however, the Wigner function is stationary under the quantum time
evolution for the stationary wave function, which is $\psi (\vec{r}%
,t\,)=\psi _{0}(\vec{r}\,)e^{-iEt/\hbar }$. Wigner function was used in
various applications and its properties investigated\cite
{muga,sala,balazs,heller,scully,bonci,berry,bonasera,royer,bos5,bos6,bos7,wig1,wig2,dos1}%
, but its true significance is when connection with classical dynamics is
sought. It is achieved by the standard assumption that classical dynamics is
the limit of quantum when $\hbar \rightarrow 0$. It can be shown in general
(one dimensional problem is discussed for the moment) that the Wigner
function satisfies the equation \cite{wig1,wig2} 
\[
\frac{\partial \rho }{\partial t}+\frac{p}{m}\frac{\partial \rho }{\partial x%
}+\frac{\partial V}{\partial x}\frac{\partial \rho }{\partial p}=-\frac{1}{2i%
}\int_{-\infty }^{\infty }dye^{ipy}G(x,y)\psi ^{\ast }(x-\hbar y/2,t)\psi
(x+\hbar y/2,t) 
\]
where 
\[
G(x,y)=\sum_{n=1}^{\infty }\frac{V^{(2n+1)}(x)}{2^{2n}(2n+1)!}(\hbar
y)^{2n+1} 
\]
and in this limit the inhomogeneous term vanishes. It also vanishes if the
potential is quadratic, irrespective of this limit. The homogeneous equation
is the Liouville equation, which determines time evolution of the phase
space density $\rho (\vec{r},\vec{p},t\,)$, and it is solved by classical
equations of motion.

Despite this, what appears, very important connection between quantum and
classical dynamics, there are at least two remarks that one can make about
this approach. It is desirable from understanding quantum-classical
connection to learn more about foundations of quantum principles because
with classical we are quite familiar. Strictly speaking that goal was not
achieved because wether the Wigner or Moyal formulation of the phase space
nothing can be learned about quantum principles because they are the
starting point anyway. The second remark concerns the limit of the phase
space density when $\hbar \rightarrow 0$, which should also be taken and not
only the limit for the inhomogeneous term. For the stationary states it can
be shown that\cite{berry} 
\[
\rho (\vec{r},\vec{p})\;%
%TCIMACRO{\underset{\hbar \rightarrow 0}{=}}%
%BeginExpansion
\mathrel{\mathop{=}\limits_{\hbar \rightarrow 0}}%
%EndExpansion
\delta \left( H-E_{0}\right) =\delta \left[ \frac{p^{2}}{2m}+V(r)-E_{0}%
\right] 
\]
where $E_{0}$ is a fixed energy. This form greatly restricts possible phase
space densities from among those stationary solutions that are obtained from
the Liouville equation. However, it is consistent with the accepted view
which is based on the correspondence principle: quantum goes over to
classical solution for large quantum numbers (the proof of this is
straightforward but not elaborated).

Classical-quantum connection can be solved by starting from entirely
classical principles\cite{bos3,bos1}, and as it will be shown allows much
greater flexibility in application of classical dynamics to quantum
problems. One starts from the Liouville equation in classical dynamics, and
the argument on which one basis its use is quite straightforward: initial
conditions for a particle are never accurately determined, in which case its
precise trajectory has no meaning, only probabilities where it is found in
the course of time. For example, to claim that position of Earth is
accurately determined is nonsense, and therefore prediction of its position
accurately in the next, say, million years is not possible, especially if
perturbation from other planets is taken into account. However, it is
possible to predict probability of finding it at certain position after
(almost) any length of time. Therefore one starts by formulating classical
mechanics as a statistical theory (it should be strongly emphasized that 
{\it statistical} does not imply {\it many} particles but probability for a
single particle), in contrast to the attempt by Moyal whose principal aim
was to formulate quantum mechanics as a statistical theory. In other words,
one particle in the phase space is not a point but an extended density,
which takes into account all uncertainties in determining its whereabouts.
For a particle of mass $m$ this equation is 
\begin{equation}
\partial _{t}\rho (\vec{r},\vec{p},t)+\frac{\vec{p}}{m}\cdot \nabla _{r}\rho
(\vec{r},\vec{p},t)+\vec{F}\cdot \nabla _{p}\rho (\vec{r},\vec{p},t)=0
\label{li}
\end{equation}
where $\vec{F}$ is the force on the particle and $\vec{p}$ is its momentum.
The index of the operator nabla designates the variable with respect to
which the derivatives are taken. For a particular case when phase space
density is stationary, which means that $\partial _{t}\rho (\vec{r},\vec{p}%
,t)=0$, the solution is a function of the form $\rho (\vec{r},\vec{p}%
,t)=f(h_{1},h_{2},h_{3},...)$, where $f$ is arbitrary function and $h_{i}$
are dynamic invariants of the classical equations of motion. One of them is
Hamiltonian for the particle, and another are components of the angular
momentum if the force $\vec{F}$ is centrally symmetric. There are other
invariants but they will be mentioned later. These phase space densities,
like any other solution of the Liouville equation, lack a very important
ingredient in order to be regarded also quantum solutions. This ingredient
is restriction on the possible phase space densities, and it is in the form
of the uncertainty principle 
\[
\Delta x\;\Delta p_{x}\succeq \hbar /2 
\]
for any Cartesian coordinate. The standard deviations $\Delta x$ and $\Delta
p$ are calculated from the phase space density. Mathematically speaking,
selecting phase space densities according to this restriction is a very well
defined problem and it is solved within the Fourier analysis. They are
obtained in the form of the convolution 
\begin{equation}
\rho (\vec{r},\vec{p},t\,)\;=\;\frac{1}{\pi ^{3}\hbar ^{3}}\int d^{3}q\,e^{2i%
\vec{p}\cdot \vec{q}/\hbar }\,f^{\ast }(\vec{r}+\vec{q}\,,t)f(\vec{r}-\vec{q}%
,t\,)  \label{roqv}
\end{equation}
which is also recognized as the Wigner function. The function $f$ is
arbitrary, but if the phase space density satisfies the Liouville equation
then it satisfies the equation\cite{bosg} 
\begin{equation}
i\hbar \;\partial _{t}f=-\frac{\hbar ^{2}}{2m}\Delta f+V\;f  \label{schr}
\end{equation}
where $\vec{F}=-\nabla V$ , and for potential it is not required to be
harmonic (for a general potential parametrization of the phase space density
is more elaborate).

The common point between the Wigner and Moyal approach and the one that
starts from classical principles is the function (\ref{wign}) or (\ref{roqv}%
), however it is almost arbitrary (because there is no reason to choose it
from many other functions) function in the former case but it is convolution
in the latter. Apart from that common point there are fundamental
differences between the two approaches, the former starts from quantum
principles and the latter from classical. One manifestation of that is
attitude towards the uncertainty relationship, which in the former case it
is not considered a principle because it is derived from the other ones,
while in the latter case it is assumed to be a fundamental principle from
which the basic dynamics equation of quantum mechanics (\ref{schr}) is
derived. One could cite other fundamental differences between the two
approaches, but one is that of simplicity. In the Wigner-Moyal approach one
needs all postulates as in formulation of quantum mechanics, about whose
number there is no general consensus, but in the approach from classical
mechanics one needs only the postulates that defines it (i.e. there is no
need for postulates other than the ones that one is already familiar width)
and the uncertainty postulate. One could question the meaning of the
uncertainty principle, but one could equally question the meaning of the
observable-operator, or wave-particle dualism postulates in the formulation
of quantum mechanics. Other advantages that one has by starting from the
classical principles and not from the quantum phase space formulation of
Wigner and Moyal, will be explicitly manifested in the following sections.
In particular it is not clear how to derive the phase space density for the
true rigid rotor, which is done in this work, by other than starting from
the classical principles.

There is one conceptual problem in the attempt to merge classical dynamics
with the uncertainty principle. Without the latter solutions of the
Liouville equation are the phase space probability densities, and as such
they always have positive value. As soon as one imposes restriction in the
form of the uncertainty principle this feature is lost and one often ends
with the phase space density that has positive and negative values. This is
the price it is paid by imposing that restriction, and its physics is
justified on the grounds that the uncertainty principle makes it impossible
to measure precisely the phase space probability density. One talks then
about the phase space density, whose properties are exactly the same as for
the probability density. However, all measurable quantities, e.g.
probability density for the coordinates only, must have physically
acceptable value.

In the equation (\ref{schr}) one recognizes Schroedinger equation, the basic
equation of quantum mechanics. Therefore the described steps in classical
mechanics produce identical results as quantum mechanics, provided the
initial condition for the phase space density is calculated from the
parameterization (\ref{roqv}). From these principles angular momentum for
the three dimensional harmonic oscillator will be analyzed first.

\section{Harmonic oscillator in three dimensions}

Stationary phase space densities for a three dimensional harmonic oscillator
will be analyzed if it is assumed that the average 
\begin{equation}
\vec{L}=\int d^{3}r\;d^{3}p\;\vec{r}\times \vec{p}\;\rho (\vec{r},\vec{p}%
\,)\;  \label{lclass}
\end{equation}
has a given value. The starting point is the phase space density which is
parametrized as (\ref{roqv}), where the functions $\psi $ are chosen in the
form $\psi =r^{l}Y_{l,\mu }(\theta ,\phi )\;R_{n,l}(r)=Y_{l,\mu
}(x,y,z)\;R_{n,l}(r)$, where $\theta $ and $\phi $ are spherical angles and $%
r$ is the radial coordinate. The indices $n,l$ and $\mu $ are integers (in
further analysis the units are set in which $m=\hbar =1$ and the frequency
of the oscillator is $\omega =1$), $Y_{l,\mu }(\theta ,\phi )$ is spherical
harmonic and 
\[
R_{n,l}(r)=N\;_{1}F_{1}\left( -n,l+\frac{3}{2}\;;\;r^{2}\right) e^{-\frac{1}{%
2}r^{2}} 
\]
where $N$ is normalization constant and $_{1}F_{1}\left( a,b;z\right) $ is
hypergeometric function. The phase space density is now 
\begin{eqnarray}
\rho _{n,l,\mu }(\vec{r},\vec{p})\; &=&\;\frac{1}{\pi ^{3}}\int d^{3}q\,e^{2i%
\vec{p}\cdot \vec{q}}\,Y_{l,\mu }^{\ast
}(x+q_{x},y+q_{y},z+q_{z})\;R_{n,l}(\left| \vec{r}+\vec{q}\right| )
\label{rohar} \\
&&Y_{l,\mu }(x-q_{x},y-q_{y},z-q_{z})\;R_{n,l}(\left| \vec{r}-\vec{q}\right|
)  \nonumber
\end{eqnarray}
which in general does not have simple explicit form, but it has few nice
features. It is stationary, which means that its form does not changes with
respect to the classical time evolution. In other words, if classical
solution for the trajectory is 
\[
\vec{r}=\vec{r}_{0}\;\cos (t)+\vec{p}_{0}\;\sin (t) 
\]
then 
\[
\rho (\vec{r},\vec{p},t)=\rho _{n,l,\mu }\left[ \vec{r}\;\cos (t)-\vec{p}%
\;\sin (t),\vec{r}\;\sin (t)+\vec{p}\;\cos (t)\right] =\rho _{n,l,\mu }(\vec{%
r},\vec{p})\; 
\]
This means that the phase space density is a function of the dynamic
invariants of the harmonic oscillator. One set of these are elements of the
energy tensor ($\frac{1}{2}$ is omitted for simplicity) $%
E_{i,j}=p_{i}p_{j}+x_{i}x_{j}\;;\;i,j=1,2,3$ (the indices designate the
Cartesian components $x,\;y$ and $z$), and the other are the components of
the angular momentum $x_{i\;}p_{j}-x_{j}\;p_{i}\;;\;i\neq j$. Of course, any
other combination of these basic invariants is possible, e.g. the total
angular momentum squared. Which ones are present in the phase space density (%
\ref{rohar}) are determined by explicit calculation, and the first few for
the ground vibrational (n=0) state are given in Table I. The symbols
represent the following quantities: $E=p^{2}+r^{2},\;E_{z}=p_{z}^{2}+z^{2},%
\;L^{2}=(yp_{z}-zp_{y})^{2}+(zp_{x}-xp_{z})^{2}+(xp_{y}-yp_{x})^{2}$ and $%
L_{z}=xp_{y}-yp_{x}$. These are dynamic invariants for the harmonic
oscillator, and therefore the phase space density is indeed stationary.

From the phase space densities one can calculate the total angular momentum (%
\ref{lclass}) and its squared modulus from 
\[
L^{2}=\int d^{3}r\;d^{3}p\;\left( \vec{r}\times \vec{p}\right) ^{2}\;\rho (%
\vec{r},\vec{p}\,)\; 
\]
Their values are given in Table I and as expected angular momentum is the
same as from quantum analysis: it has only the z component and its value is $%
\mu $. However, the angular momentum squared is not equal to $l(l+1)$, as
expected from the quantum treatment, but differs by $3/2$. The same is true
for other than $n=0$ states, as shown in Table II. In fact the most
surprising finding is that the states with $l=0$, which are normally
associated with the zero angular momentum, have the value $3/2$ for the
angular momentum squared. The question is where this discrepancy comes from?
The simplest answer is that classical analysis is not correct, because it
surely must violate certain rules that are not consistent with the quantum
mechanical ones. However, this answer is not correct, because if one writes
the momentum squared as $p^{2}=p_{r}^{2}+p_{\theta }^{2}+p_{\phi }^{2}$,
where the components of the momentum are given with respect to the vector $%
\vec{r}$, then its average for the function $\psi =\Theta (\theta ,\phi
)R(r) $ is 
\begin{eqnarray*}
&<&p^{2}>=\frac{1}{\pi ^{3}}\int d^{3}r\;dp_{r}\;dp_{\theta }\;dp_{\phi
}\;\left( p_{r}^{2}+p_{\theta }^{2}+p_{\phi }^{2}\right) \\
&&\int dq_{r}\;dq_{\theta }\;dq_{\phi }\;e^{2i\left( p_{r}q_{r}+p_{\theta
}q_{\theta }+p_{\phi }q_{\phi }\right) }\Theta ^{\ast }(\theta _{+},\phi
_{+})R(r_{+})\Theta (\theta _{-},\phi _{-})R(r_{-})
\end{eqnarray*}
where 
\begin{eqnarray}
r_{\pm } &=&\sqrt{(r\pm q_{r})^{2}+q_{\theta }^{2}+q_{\phi }^{2}\,}%
\;\;;\;\;\cos \theta _{\pm }=\frac{\hat{z}\cdot (\vec{r}\pm \vec{q}\,)}{|%
\vec{r}\pm \vec{q}\,|}\;=\;\frac{(r\pm q_{r})\;\cos \theta \pm q_{\theta
}\,\;\sin \theta }{\sqrt{(r\pm q_{r})^{2}+q_{\theta }^{2}+q_{\phi }^{2}\,}}
\label{rthph} \\
e^{i\phi _{\pm }} &=&\frac{(\vec{r}\pm \vec{q}\,)\cdot \left( \hat{x}+i\hat{y%
}\right) }{|\vec{r}\pm \vec{q}\,|\sin \theta _{\pm }}=\frac{(r\pm q_{r})\sin
\theta \mp q_{\theta }\cos \theta \pm iq_{\phi }}{|\vec{r}\pm \vec{q}\,|\sin
\theta _{\pm }}e^{i\phi }  \nonumber
\end{eqnarray}
After a straightforward, but lengthy, calculation of the integrals one
obtains 
\[
<p^{2}>=-\int d^{3}r\;\psi ^{\ast }(\vec{r})\frac{1}{r^{2}}\left[ \frac{d}{dr%
}\left( r^{2}\frac{d}{dr}\right) \,+\frac{1}{sin(\theta )}\frac{\partial }{%
\partial \theta }\left( sin(\theta )\frac{\partial }{\partial \theta }%
\right) \,+\,\frac{1}{sin^{2}(\theta )}\frac{\partial ^{2}}{\partial \phi
^{2}}\right] \psi (\vec{r}) 
\]
which is the correct answer for the kinetic energy operator (up to a
pre-factor, which was not taken into account). In the standard
interpretation the angular part is then associated with the angular momentum
squared operator, which indeed gives zero for angular momentum squared for
the $l=0$ states. Therefore the procedure of deriving the kinetic energy
operator is correct, but then the question is where the analysis that
produced results in Tables I and II is inconsistent with the quantum
interpretation? To answer this question one explicitly calculates the
average of $p_{\Omega }^{2}=p_{\theta }^{2}+p_{\phi }^{2}$, which is the
classical angular part of the momentum squared. For simplicity its average
will be calculated for $l=0$ state.

By definition 
\begin{eqnarray*}
&<&p_{\Omega }^{2}>=\frac{1}{4\pi ^{4}}\int d^{3}r\;dp_{r}\;dp_{\theta
}\;dp_{\phi }\;\left( p_{\theta }^{2}+p_{\phi }^{2}\right) \\
&&\int dq_{r}\;dq_{\theta }\;dq_{\phi }\;e^{2i\left( p_{r}q_{r}+p_{\theta
}q_{\theta }+p_{\phi }q_{\phi }\right) }R(\sqrt{(r+q_{r})^{2}+q_{\theta
}^{2}+q_{\phi }^{2}\,})R(\sqrt{(r-q_{r})^{2}+q_{\theta }^{2}+q_{\phi }^{2}\,}%
)
\end{eqnarray*}
where integration in $p_{r}$ and $q_{r}$ is calculated first, then $%
p_{\theta }^{2}$ and $p_{\phi }^{2}$ are replaced by derivatives in the
variables $q_{\theta }$ and $q_{\phi }$ that act on the exponential
function, respectively, and after partial integration in the same variables
one obtains 
\[
<p_{\Omega }^{2}>=-\frac{1}{16\pi ^{3}}\int d^{3}r\;dp_{\theta }\;dp_{\phi
}\;\int dq_{\theta }\;dq_{\phi }\;e^{2i\left( p_{\theta }q_{\theta }+p_{\phi
}q_{\phi }\right) }\left( \partial _{q_{\theta }}^{2}+\partial _{q_{\phi
}}^{2}\right) \left[ R^{2}(\sqrt{r^{2}+q_{\theta }^{2}+q_{\phi }^{2}\,})%
\right] 
\]
After integration in the variables $p_{\theta }\;$and $p_{\phi }$ the only
non zero contribution is 
\[
<p_{\Omega }^{2}>=-\frac{1}{4\pi }\int d^{3}r\;R(r)\;\frac{1}{r}\partial
_{r}R(r) 
\]
and it is not zero. In fact the average of the angular momentum squared
(which is $L^{2}=r^{2}p_{\Omega }^{2}$) is 
\[
<L^{2}>=-\int_{0}^{\infty }dr\;r^{3}R(r)\;\partial _{r}R(r)=\frac{3}{2} 
\]
which is precisely the value in Tables I and II. Furthermore the result is
independent of the radial function, which indicates that this is a universal
number. In order to check the correctness of the result one calculates the
average of the radial part of the momentum squared, and the result is 
\[
<p_{r}^{2}>=-\frac{1}{4\pi }\int d^{3}r\;R(r)\;\left[ \partial _{r}^{2}R(r)+%
\frac{1}{r}\partial _{r}R(r)\right] 
\]
Together with the angular part one obtains the radial part of the kinetic
energy operator 
\[
T_{r}=-\frac{1}{r^{2}}\partial _{r}\left( r^{2}\partial _{r}\right) 
\]
which is the correct answer. Therefore, there is no discrepancy between
quantum expression for the kinetic energy and classical treatment in this
work, but the difference is in the interpretation of its fragments.
According to results in this work one should interpret the radial kinetic
energy operator as 
\[
T_{r}=-\partial _{r}^{2}-\frac{1}{r}\partial _{r} 
\]
while 
\[
T_{\Omega }=-\frac{1}{r}\partial _{r}-\frac{1}{r^{2}}\left[ \frac{1}{%
sin(\theta )}\frac{\partial }{\partial \theta }\left( sin(\theta )\frac{%
\partial }{\partial \theta }\right) \,+\,\frac{1}{sin^{2}(\theta )}\frac{%
\partial ^{2}}{\partial \phi ^{2}}\right] 
\]
is operator for the angular momentum squared.

Further confirmation of this comes from calculating the average angular
momentum explicitly. For $l=0$ states it is given by 
\[
<\vec{L}>=<L_{z}>\hat{z}=\frac{\hat{z}}{4\pi ^{4}}\int
d^{3}r\;dp_{r}\;dp_{\theta }\;dp_{\phi }\;r\;p_{\phi }\;\sin (\theta )\int
dq_{r}\;dq_{\theta }\;dq_{\phi }\;e^{2i\left( p_{r}q_{r}+p_{\theta
}q_{\theta }+p_{\phi }q_{\phi }\right) }R(r_{+})R(r_{-}) 
\]
which is easily shown to be zero. This is no contradiction with the previous
finding, because zero of the angular momentum is result of cancellations of
the contributions from two signs of $p_{\phi }$ rather than the modulus of
the angular momentum being zero. One can therefore calculate angular
momentum by averaging over only one sign of $p_{\phi }$, because the average
over the other gives the same result but with the opposite sign, and the
sum-total is zero. This average is 
\[
<L_{z}^{+}>=\frac{1}{4\pi ^{4}}\int d^{3}r\;dp_{r}\;dp_{\theta }\;\sin
(\theta )\int_{0}^{\infty }dp_{\phi }\;r\;p_{\phi }\int dq_{r}\;dq_{\theta
}\;dq_{\phi }\;e^{2i\left( p_{r}q_{r}+p_{\theta }q_{\theta }+p_{\phi
}q_{\phi }\right) }R(r_{+})R(r_{-}) 
\]
and after evaluating most of integrals one gets 
\[
<L_{z}^{+}>=-\frac{1}{8i\pi ^{2}}\int d^{3}r\;r\;\sin (\theta
)\int_{0}^{\infty }dp_{\phi }\int dq_{\phi }\;e^{2ip_{\phi }q_{\phi
}}\partial _{q_{\phi }}R^{2}(\sqrt{r^{2}+q_{\phi }^{2}\,}) 
\]
By using the relationship 
\[
\int_{0}^{\infty }dp_{\phi }\;e^{2ip_{\phi }q_{\phi }}=\pi \delta (2q_{\phi
})+i\;\P \left( \frac{1}{2q_{\phi }}\right) 
\]
where \P\ designates the principal value of the integral, it is obtained 
\[
<L_{z}^{+}>=-\frac{i}{8i\pi ^{2}}\int d^{3}r\;r\sin (\theta )\;\P \left[
\int dq_{\phi }\;R(\sqrt{r^{2}+q_{\phi }^{2}\,})R^{\prime }(\sqrt{%
r^{2}+q_{\phi }^{2}\,})\frac{1}{\sqrt{r^{2}+q_{\phi }^{2}\,}}\right] 
\]
The principal value can be omitted because the integrand is not singular, in
which case 
\[
<L_{z}^{+}>=\frac{1}{4}\int dr\;r^{2}\;R^{2}(r)=\frac{1}{4} 
\]
The value of the z component of the angular momentum, which gets
contribution from the space $p_{\phi }>0$, is independent of the radial
function $R(r)$, the same result as before.

\section{Rigid rotor}

Rigid rotor in classical physics is a very well defined object, but one
could argue that it is also in quantum. Unfortunately, as it will be shown,
the two definitions have not the same meaning. Intuition, which is
essentially classical, defines rigid object as the one which does not have
breathing (radial) motion. Objects that appear to qualify as the rigid are
diatomic molecules (at least many of them), which is supported by the
approximations in their quantum description. Schroedinger equation for the
nuclear motion in the diatomic molecule is 
\[
i\hbar \frac{\partial \psi }{\partial t}=-\frac{\hbar ^{2}}{2m}\left[ \frac{1%
}{r^{2}}\frac{\partial }{\partial r}\left( r^{2}\frac{\partial \psi }{%
\partial r}\right) +\frac{1}{r^{2}sin(\theta )}\frac{\partial }{\partial
\theta }\left( sin(\theta )\frac{\partial \psi }{\partial \theta }\right)
\,+\,\frac{1}{r^{2}sin^{2}(\theta )}\frac{\partial ^{2}\psi }{\partial \phi
^{2}}\right] +V(r)\;\psi 
\]
where $m$ is the reduced mass of the diatomic molecule and $V(r)$ is the
internuclear potential. Typically this potential has a deep minimum
displaced from the origin by the distance $r_{0}$ (the bond length), which
for the modelling purpose is approximated by $\frac{m}{2}\omega
^{2}(r-r_{0})^{2}$. Ground state wave function in this potential has the
width $a$, and its relationship to the bond length is $r_{0}\gg a$. Under
this circumstance the radial coordinate in the angular part of the kinetic
energy operator is approximated by the fixed value $r_{0}$, in which case
the wave function factors as $\psi =\frac{1}{r}R(r)\Theta (\theta ,\phi
,t)e^{-iE_{0}t/\hbar }$, where $E_{0}$ is the ground state energy of the
molecule. The angular function then satisfies the equation 
\begin{equation}
i\hbar \frac{\partial \Theta (\theta ,\phi ,t)}{\partial t}=T_{rot}\;\Theta
(\theta ,\phi ,t)+W(\theta ,\phi ,t)\Theta (\theta ,\phi ,t)  \label{roteq}
\end{equation}
where 
\[
T_{rot}=-\frac{\hbar ^{2}}{2mr_{0}^{2}}\left[ \frac{1}{sin(\theta )}\frac{%
\partial }{\partial \theta }\left( sin(\theta )\frac{\partial }{\partial
\theta }\right) \,+\,\frac{1}{sin^{2}(\theta )}\frac{\partial ^{2}}{\partial
\phi ^{2}}\right] 
\]
and $W(\theta ,\phi ,t)$ represents interaction of the molecule that depends
on the angles only (say a dipole in the electric field).

The unique feature of the equation (\ref{roteq}) is that it depends only on
the angle variables, orientations angles of the molecule, and in this sense
it is called the equation for the rigid rotor. The operator $T_{rot}$ then
represents the kinetic energy operator for the rigid rotor, which has
discrete spectrum 
\[
T_{rot}\;Y_{l,m}(\theta ,\phi )=\frac{\hbar ^{2}}{2mr_{0}^{2}}%
l(l+1)\;Y_{l,m}(\theta ,\phi
)\;\;\;\;;\;\;\;l=0,1,2,..\;\;\;\;m=-l,-l+1,....,l 
\]
where $Y_{l,m}(\theta ,\phi )$ are spherical harmonics, the eigenfunctions
of the rigid rotor. However, this is far from being true, if one has the
image of the classical rigid rotor in mind. Although in the previous
derivation dynamics in the angle variables is de-coupled from the dynamics
in the radial this is not sufficient to call the molecule a rigid rotor. It
is also necessary to prove that the radial component of the momentum for the
relative motion of the two atoms is de-coupled from its angular ones, and to
show this one needs to calculate the momentum space wave function. This
function for a stationary rotational state of the ''rigid rotor'' is 
\begin{equation}
\varphi (\vec{p})=\int d^{3}r\;\frac{1}{r}R(r)Y_{l,m}(\theta ,\phi )\;e^{i%
\vec{p}\cdot \vec{r}}=\left( 2\pi \right) ^{3/2}i^{l}Y_{l,m}(\theta
_{p},\phi _{p})\frac{1}{\sqrt{p}}\int_{0}^{\infty
}dr\;r^{1/2}R(r)J_{l+1/2}(pr)  \label{mom}
\end{equation}
where the expansion of the plane wave in the spherical coordinates was used.
De-coupling of the radial component $p_{r}$ from its angular ones $p_{\theta
}$ and $p_{\phi }$ is therefore not possible because $p=\sqrt{%
p_{r}^{2}+p_{\theta }^{2}+p_{\phi }^{2}}$, while the spherical angles $%
\theta _{p}$ and $\phi _{p}$ are related to the same components in a
complicated way. Therefore it cannot be assumed that the theoretical model
that was just described represents rigid rotor.

Analysis in the phase space is more revealing, but it is somewhat
approximate. It was mentioned that around $r=r_{0}$ the potential is very
well approximated by harmonic, in which case the entire wave function for
the diatomic molecule is 
\[
\phi _{l}(\vec{r})\;=N\;e^{-\frac{(r-r_{0})^{2}}{2a^{2}}}Y_{l,m}(\theta
,\phi ) 
\]
where $1/r$ coefficient was assumed to be constant. The phase space density
is then 
\[
\rho _{l,m}(\vec{r},\vec{p})\;=\;\frac{N^{2}}{\pi ^{3}}\int d^{3}q\,e^{2i%
\vec{p}\cdot \vec{q}}\,Y_{l,m}^{\ast }(\theta _{+},\phi
_{+})\;Y_{l,m}(\theta _{-},\phi _{-})e^{-\frac{(\left| \vec{r}+\vec{q}%
\right| -r_{0})^{2}}{2a^{2}}-\frac{(\left| \vec{r}-\vec{q}\right| -r_{0})^{2}%
}{2a^{2}}} 
\]
where the angles $\theta _{\pm }$ and $\phi _{\pm }$ are defined in (\ref
{rthph}). The rigid rotor model assumption implies that $q<<r$, and if
spherical coordinates are used for the vector $\vec{q}=$ $q_{r}\hat{r}%
+q_{\theta }\hat{\theta}+q_{\phi }\vec{\phi}$ then the exponent of the
harmonic oscillator function is approximately

\[
\frac{(\left| \vec{r}+\vec{q}\right| -r_{0})^{2}}{2a^{2}}+\frac{(\left| \vec{%
r}-\vec{q}\right| -r_{0})^{2}}{2a^{2}}\sim \frac{(r-r_{0})^{2}}{a^{2}}+\frac{%
q_{r}^{2}}{a^{2}}\left( 1-\frac{q_{\theta }^{2}+q_{\phi }^{2}}{r_{0}^{2}}%
\right) +\frac{\left( q_{\theta }^{2}+q_{\phi }^{2}\right) ^{2}}{%
4a^{2}r_{0}^{2}}+O(q^{6}) 
\]
where in the expansion coefficients, except in the leading one, it was set $%
r=r_{0}$ . By recalling that this expansion is in the exponent it follows
that the range of $\left| q_{r}\right| $ within which it significantly
contributes to the phase space density is of the order $a$, however, the
range of $\sqrt{q_{\theta }^{2}+q_{\phi }^{2}}$ is of the order $\sqrt{ar_{0}%
}$. This means that $\left| q_{r}\right| <<\sqrt{q_{\theta }^{2}+q_{\phi
}^{2}}$ in which case the phase space density is approximately 
\begin{eqnarray*}
\rho _{l,m}(\vec{r},\vec{p})\; &=&\frac{2r_{0}a^{2}N^{2}}{\pi ^{5/2}}e^{-%
\frac{(r-r_{0})^{2}}{a^{2}}-p_{r}^{2}a^{2}} \\
&&\int_{-\infty }^{\infty }dq_{\theta }\int_{-\infty }^{\infty }dq_{\phi
}\,\;e^{2i\sqrt{2ar_{0}}\left( p_{\theta }q_{\theta }+p_{\phi }q_{\phi
}\right) -\left( q_{\theta }^{2}+q_{\phi }^{2}\right) ^{2}}\;\Theta
_{l,m}(r_{0},\theta ,\phi ,0,\sqrt{2ar_{0}}q_{\theta },\sqrt{2ar_{0}}q_{\phi
})
\end{eqnarray*}
where the angular function $\Theta _{l,m}$ is shorthand for the product of
two functions $Y_{l,m}(\theta _{\pm },\phi _{\pm })$. Dependence of the
angular function on the radial variable $q_{r}$ is also neglected, in which
case the phase space density parametrizes as a product of the function for
the radial variables and the function for the angular. The angular phase
space density is now defined as 
\begin{equation}
\rho _{l,m}^{\Omega }(\theta ,\phi ,p_{\theta },p_{\phi })\;=\frac{2ar_{0}}{%
\pi ^{2}}\int dq_{\theta }\;dq_{\phi }\,\;e^{2i\sqrt{2ar_{0}}\left(
p_{\theta }q_{\theta }+p_{\phi }q_{\phi }\right) -\left( q_{\theta
}^{2}+q_{\phi }^{2}\right) ^{2}}\;\Theta _{l,m}(r_{0},\theta ,\phi ,0,\sqrt{%
2ar_{0}}q_{\theta },\sqrt{2ar_{0}}q_{\phi })  \label{3dphas}
\end{equation}
and appears to describe the rigid rotor because it is independent of the
radial variables. However, this conclusion is false because the phase space
density depends on the radial parameter $a$, and in this respect it depends
on the radial dynamics. Because of this feature the rigid rotor will be
called ''soft''.

Average of the angular momentum squared has now the value 
\[
<L^{2}>=\int \rho _{l,m}^{\Omega }(\theta ,\phi ,p_{\theta },p_{\phi
})r_{0}^{2}\left( p_{\theta }^{2}+p_{\phi }^{2}\right) =l(l+1) 
\]
where the constant term $3/2$ that was obtained in the previous section is
missing. This is expected because coupling with the radial motion was
neglected. Another property of the angular phase space density is 
\begin{equation}
\int_{-\infty }^{\infty }dp_{\theta }\int_{-\infty }^{\infty }dp_{\phi
}\,\;\rho _{l,m}^{\Omega }(\theta ,\phi ,p_{\theta },p_{\phi })=\Theta
_{l,m}(r_{0},\theta ,\phi ,0,0,0)=\;\left| Y_{l,m}(\theta ,\phi )\right| ^{2}
\label{sphhar}
\end{equation}
which shows that it correctly describes the essentials of the angular
momentum. Its explicit form for $l=m=0$ is given by 
\[
\rho _{0,0}^{\Omega }(\theta ,\phi ,p_{\theta },p_{\phi })\;=\frac{ar_{0}}{%
\pi ^{2}}\int_{0}^{\infty }dq\;qJ_{0}\left( 2\sqrt{2ar_{0}}pq\right)
e^{-q^{4}} 
\]
which does not have explicit analytic expression, but its shape is simple,
as shown in Figure 1. For the rotational state $l=1$ and $m=0$ the angular
phase space density is 
\begin{eqnarray*}
\rho _{1,0}^{\Omega }(\theta ,\phi ,p_{\theta },p_{\phi })\; &=&\frac{6ar_{0}%
}{4\pi ^{3}}\int_{-\infty }^{\infty }dq_{\theta }\int_{-\infty }^{\infty
}dq_{\phi }\,\;e^{2i\sqrt{2ar_{0}}\left( p_{\theta }q_{\theta }+p_{\phi
}q_{\phi }\right) -\left( q_{\theta }^{2}+q_{\phi }^{2}\right) ^{2}}\;\;%
\frac{\cos ^{2}\theta -2\epsilon q_{\theta }^{2}\,\;\sin ^{2}\theta }{%
1+2\epsilon q_{\theta }^{2}+2\epsilon q_{\phi }^{2}\,} \\
&\sim &\left( \cos ^{2}\theta +\frac{1}{4r_{0}^{2}}\partial _{p_{\theta
}}^{2}+\frac{\cos ^{2}\theta }{4r_{0}^{2}}\partial _{p_{\phi }}^{2}\right)
\int_{0}^{\infty }dq\;qJ_{0}\left( 2\sqrt{2ar_{0}}pq\right) e^{-q^{4}}\;\;
\end{eqnarray*}
where $\epsilon =a/r_{0}$, and in the last step only the terms up to the
order $\epsilon $ where retained. Similarly the phase space density for $l=1$
and $m=1$ is 
\[
\rho _{1,1}^{\Omega }(\theta ,\phi ,p_{\theta },p_{\phi })\sim \left( \sin
^{2}\theta +\frac{\sin \theta }{r_{0}}\partial _{p_{\phi }}+\frac{1}{%
4r_{0}^{2}}\partial _{p_{\theta }}^{2}+\frac{1}{4r_{0}^{2}}\partial
_{p_{\phi }}^{2}+\frac{\sin ^{2}\theta }{4r_{0}^{2}}\partial _{p_{\phi
}}^{2}\right) \int_{0}^{\infty }dq\;qJ_{0}\left( 2\sqrt{2ar_{0}}pq\right)
e^{-q^{4}}\;\; 
\]

These phase space densities, except $\rho _{0,0}^{\Omega }$, are time
dependent, which means that they are not functions of only the dynamic
invariants of the rigid rotor, e.g. $p^{2}=p_{\theta }^{2}+p_{\phi }^{2}$
and $p_{\phi }\sin \theta $. Thus for example the terms in $\rho
_{1,0}^{\Omega }(\theta ,\phi ,p_{\theta },p_{\phi })$ that cannot be
represented by invariants are 
\[
\rho _{1,0}^{\Omega }(\theta ,\phi ,p_{\theta },p_{\phi })\;=\cos ^{2}\theta
\int_{0}^{\infty }dq\;qJ_{0}\left( 2\sqrt{2ar_{0}}pq\right) e^{-q^{4}}-\frac{%
\cos ^{2}\theta }{pr_{0}^{2}}\sqrt{\frac{ar_{0}}{2}}\int_{0}^{\infty
}dq\;q^{2}J_{1}\left( 2\sqrt{2ar_{0}}pq\right) e^{-q^{4}} 
\]
Explicit time dependence of the phase space density is obtained from the
time dependence of the angle $\theta $ 
\[
\cos \theta =\cos \theta _{0}\;\cos \frac{tp_{0}}{mr_{0}}-\frac{p_{\theta
}^{0}}{p_{0}}\sin \theta _{0}\;\sin \frac{tp_{0}}{mr_{0}} 
\]
where $\theta _{0}$, $p_{\theta }^{0}$ and $p_{0}$ are initial values of
these variables. In the phase space density $\cos \theta $ is replaced by 
\[
\cos \theta _{t}=\cos \theta \;\cos \frac{tp}{mr_{0}}+\frac{p_{\theta }}{p}%
\sin \theta \;\sin \frac{tp}{mr_{0}} 
\]
and the time dependence of the probability density (\ref{sphhar}) is
calculated from 
\begin{gather*}
\int_{-\infty }^{\infty }dp_{\theta }\int_{-\infty }^{\infty }dp_{\phi
}\,\;\rho _{1,0}^{\Omega }(\theta ,\phi ,p_{\theta },p_{\phi })\sim
\int_{-\infty }^{\infty }dp_{\theta }\int_{-\infty }^{\infty }dp_{\phi
}\,\;\cos ^{2}\theta \int_{0}^{\infty }dq\;qJ_{0}\left( 2\sqrt{2ar_{0}}%
pq\right) e^{-q^{4}} \\
=\frac{\pi }{32ar_{0}}\left( 1+\cos ^{2}\theta \right) \;-\frac{\pi mr_{0}}{8%
\sqrt{2ar_{0}}}\left( 3\cos ^{2}\theta -1\right) \partial _{t}\int_{0}^{%
\frac{t}{mr_{0}\sqrt{2ar_{0}}}}dq\;\frac{qe^{-q^{4}}}{\sqrt{\frac{t^{2}}{%
2am^{2}r_{0}^{3}}-q^{2}}}
\end{gather*}
The first part is time independent, while the second goes to zero after the
time interval $t>mr_{0}\sqrt{2ar_{0}}$, which is typically of the order $%
10^{-13}-10^{-14}\;\sec $ for diatomic molecules. Therefore the probability
density starts as (\ref{sphhar}) but its limiting value is constant, but not
in the form of the squared modulus of the spherical harmonic.

\section{True rigid rotor}

Previous discussion revealed great difficulty in formulating the concept of
the rigid rotor in quantum theory. This was manifested as inability to
formulate the phase space density that involves only the parameters for the
rotational degrees of freedom. It does not help to take the limit $%
a\longrightarrow 0$ in (\ref{3dphas}) because that would imply infinite
dispersion of the variables $p_{\theta }$ and $p_{\phi }$, which only
reflects the fact that the radial and the angular components of the momentum
are interrelated. There is, however, a way of formulating the true rigid
rotor, but should be done by following formulation of quantum mechanics as
suggested in Introduction. One starts from the Liouville equation in the
spherical coordinates, which for a free particle is

\begin{eqnarray*}
&&\partial _{t}\rho -\frac{p_{\theta }}{mr}\partial _{\theta }\rho +\frac{%
p_{\phi }}{mr\,\sin \theta }\partial _{\phi }\rho +\frac{p_{r}}{m}\partial
_{r}\rho - \\
&&\frac{1}{mr}\,\left( \frac{p_{\phi }^{2}\cos \theta }{\,\sin \theta }%
\,+p_{r}\,p_{\theta }\right) \;\partial _{p_{\theta }}\rho -\frac{p_{\phi }}{%
mr\,}\left( -p_{\theta }\,\frac{\,\cos \theta }{\,\sin \theta }+p_{r}\right)
\;\partial _{p_{\phi }}\rho \;+\,\left( \,\frac{p_{\theta }^{2}}{mr}+\,\frac{%
p_{\phi }^{2}}{mr\,}\right) \;\partial _{p_{r}}\rho =\;0
\end{eqnarray*}
and the rigid rotor assumption implies that the phase space density is $r$
and $p_{r}$ independent. This means that the Liouville equation for the
rigid rotor is 
\begin{equation}
\partial _{t}\rho -\frac{p_{\theta }}{mr}\partial _{\theta }\rho +\frac{%
p_{\phi }}{mr\,\sin \theta }\partial _{\phi }\rho +\frac{p_{\phi }}{mr\,}\,%
\frac{\,\cos \theta }{\,\sin \theta }\left( p_{\theta }\;\partial _{p_{\phi
}}\rho -p_{\phi }\;\partial _{p_{\theta }}\rho \right) \;=\;0  \label{rotli}
\end{equation}
where $r$ is constant. It can be easily verified that if the phase space
density is a function of the form $\rho (\theta ,\phi ,p_{\theta },p_{\phi
},t)=F(p_{\theta }^{2}+\,p_{\phi }^{2},p_{\phi }\sin \theta )$, where $F$ is
arbitrary function, then it is time independent. Additional requirement is
that the phase space density should be in accordance with the uncertainty
principle, which is achieved by straightforward generalization of the rule
that was used before. The phase space density is therefore parametrized as 
\begin{equation}
\rho (\theta ,\phi ,p_{\theta },p_{\phi },t)\;=\;\frac{1}{\pi ^{2}}\int
d^{2}q\,e^{2i\left( p_{\theta }q_{\theta }+p_{\phi }q_{\phi }\right)
}\,\Theta ^{\ast }(\theta _{+},\phi _{+})\;\Theta (\theta _{-},\phi _{-})
\label{rorig}
\end{equation}
where the angular functions will be determined for a particular case when
the solutions of the Liouville equation (\ref{rotli}) are stationary, i.e. $%
\partial _{t}\rho =0$. The relevant variables were defined in (\ref{rthph}).
Parametrization (\ref{rorig}) is replaced in the Liouville equation (\ref
{rotli}), and by using transformations of the kind 
\begin{eqnarray*}
p_{\theta }\rho (\theta ,\phi ,p_{\theta },p_{\phi }) &=&-\frac{1}{2i}\int
d^{2}q\,e^{2i\left( p_{\theta }q_{\theta }+p_{\phi }q_{\phi }\right)
}\partial _{q_{\theta }}\left[ \Theta (\theta _{-},\phi _{-})\;\Theta ^{\ast
}(\theta _{+},\phi _{+})\right] \\
p_{\theta }\partial _{p_{\theta }}\rho (\theta ,\phi ,p_{\theta },p_{\phi })
&=&-\int d^{2}q\,e^{2i\left( p_{\theta }q_{\theta }+p_{\phi }q_{\phi
}\right) }\partial _{q_{\theta }}\left[ q_{\theta }\;\Theta (\theta
_{-},\phi _{-})\;\Theta ^{\ast }(\theta _{+},\phi _{+})\right]
\end{eqnarray*}
one obtains, after lengthy simplifications, that the stationary solutions of
the Liouville equation (\ref{rotli}) satisfy 
\begin{equation}
\frac{i}{2\pi ^{2}}\int d^{2}q\,e^{2i\left( p_{\theta }q_{\theta }+p_{\phi
}q_{\phi }\right) }\frac{1}{r^{2}+q^{2}}\left( f_{1}+\frac{1}{r}f_{2}\right)
=0  \label{stat}
\end{equation}
where 
\[
f_{1}=\,\Theta ^{\ast }\;\left[ \frac{1}{\sin \theta _{-}}\partial _{\theta
_{-}}\left( \sin \theta _{-}\;\partial _{\theta _{-}}\Theta \right) +\frac{1%
}{\sin ^{2}\theta _{-}}\partial _{\phi _{-}}^{2}\Theta \right] -\,\Theta \;%
\left[ \frac{1}{\sin \theta _{+}}\partial _{\theta _{+}}\left( \sin \theta
_{+}\;\partial _{\theta _{+}}\Theta ^{\ast }\right) +\frac{1}{\sin
^{2}\theta _{+}}\partial _{\phi +}^{2}\Theta ^{\ast }\right] \; 
\]
and 
\begin{eqnarray*}
f_{2} &=&-\frac{q^{2}}{2r}(\cos \theta _{+}+\cos \theta _{-})\left( \frac{%
\Theta ^{\ast }\partial _{\theta _{-}}\Theta }{\sin \theta _{-}}-\frac{%
\Theta \partial _{\theta _{+}}\Theta ^{\ast }}{\sin \theta _{+}}\right) -%
\frac{r}{2}(\cos \theta _{+}-\cos \theta _{-})\left( \frac{\Theta ^{\ast
}\partial _{\theta _{-}}\Theta }{\sin \theta _{-}}+\frac{\Theta \partial
_{\theta _{+}}\Theta ^{\ast }}{\sin \theta _{+}}\right) + \\
&&q_{\phi }\sin \theta \left( \frac{\Theta ^{\ast }\partial _{\phi
_{-}}\Theta }{\sin ^{2}\theta _{-}}+\frac{\Theta \partial _{\phi _{+}}\Theta
^{\ast }}{\sin ^{2}\theta _{+}}\right)
\end{eqnarray*}
where $q^{2}=q_{\theta }^{2}+q_{\phi }^{2}$. It is implied that $\Theta $ is
a function of the variables $\theta _{-}$ and $\phi _{-}$ while $\Theta
^{\ast }$ is a function of $\theta _{+}$ and $\phi _{+}$.

If the function $f_{2}$ is neglected for the moment then the condition (\ref
{stat}) implies that the angular function satisfies the differential
equation 
\[
\frac{1}{\sin \theta }\partial _{\theta }\left( \sin \theta \;\partial
_{\theta }\Theta \right) +\frac{1}{\sin ^{2}\theta }\partial _{\phi
}^{2}\Theta =\lambda \;\Theta 
\]
where $\lambda $ is a real constant. In this equation one recognizes the
equation for the spherical harmonics, where $\lambda =-l(l+1)$. However, the
function $f_{2}$ cannot be neglected, and therefore the angular functions
are only approximately the spherical harmonics. However, a very useful
feature of the function $f_{2}$ is that in the limits $q_{\theta
}\rightarrow 0$ and $q_{\phi }\rightarrow 0$ it is equal to zero, in which
case the angular functions are exactly the spherical harmonics. This means
that when the phase space density is integrated over the momentum variables
the resulting probability density should be the squared modulus of the
spherical harmonics, i.e. 
\begin{equation}
P(\theta ,\phi )=\int dp_{\theta }\;\int dp_{\phi }\;\rho (\theta ,\phi
,p_{\theta },p_{\phi })\;=\,\left| \Theta (\theta ,\phi )\right|
^{2}=\;\,\left| Y_{l,m}(\theta ,\phi )\right| ^{2}  \label{vjer}
\end{equation}
The choice of the spherical harmonics for the angular functions is that the
phase space density is approximate, which is manifested as being time
dependent, i.e. it is not a function of only the dynamic invariants of the
rigid rotor. This is the price that is paid by neglecting the function $%
f_{2} $ in the equation (\ref{stat}). Inclusion of this function results in
the phase space density that is a function of only these invariants, and
this fact is used as the procedure to find a proper phase space density.
This is best demonstrated on one example. One particular case, however, has
exact solution, and this is when the angular function is constant. In this
case 
\[
\rho _{0,0}(\theta ,\phi ,p_{\theta },p_{\phi })\;=\;\frac{1}{\pi ^{2}}\int
d^{2}q\,e^{2i\left( p_{\theta }q_{\theta }+p_{\phi }q_{\phi }\right) }\,%
\frac{1}{4\pi }=\frac{1}{4\pi }\delta (p_{\theta })\delta (p_{\phi }) 
\]

The example that will be analyzed in more details is when the angular
function is the spherical harmonic $Y_{1,0}(\theta ,\phi )$. The phase space
density (\ref{rorig}) is 
\[
\rho (\theta ,\phi ,p_{\theta },p_{\phi },t)\;=\;\frac{1}{\pi ^{2}}\int
d^{2}q\,e^{2i\left( p_{\theta }q_{\theta }+p_{\phi }q_{\phi }\right)
}\,Y_{1,0}^{\ast }(\theta _{+},\phi _{+})\;Y_{1,0}(\theta _{-},\phi _{-}) 
\]
and its explicit form is (the constant $r\;$is fixed to unity) 
\begin{eqnarray*}
\rho (\theta ,\phi ,p_{\theta },p_{\phi },t)\; &=&\;\frac{3}{4\pi ^{3}}\int
d^{2}q\,e^{2i\left( p_{\theta }q_{\theta }+p_{\phi }q_{\phi }\right) }\,%
\frac{q_{\theta }^{2}\sin ^{2}\theta -\cos ^{2}\theta }{1+q^{2}}= \\
&&\frac{3}{2\pi ^{2}}\left( 1-\frac{p_{\phi }^{2}\sin ^{2}\theta }{p^{2}}%
\right) K_{0}(2p)+\frac{3\sin ^{2}\theta }{4p\pi ^{2}}\left( 1-2\frac{%
p_{\phi }^{2}}{p^{2}}\right) K_{1}(2p)-\frac{3}{8\pi }\delta (p_{\theta
})\delta (p_{\phi })\sin ^{2}\theta
\end{eqnarray*}
where $K_{n}(x)$ is modified Bessel function of the second kind. This phase
space density is not stationary, because the term 
\[
\rho _{t}=\frac{3\sin ^{2}\theta }{4p\pi ^{2}}K_{1}(2p) 
\]
is not a combination of the dynamic invariants. However, if $p_{\phi }$ is
replaced by $p\;\cos \alpha $ then 
\[
\int_{0}^{2\pi }d\alpha \;\left( 1-2\frac{p_{\phi }^{2}}{p^{2}}\right) =0 
\]
which means that these two terms in the phase space density can be omitted
without in any way modifying the integral (\ref{vjer}), the value of the
total angular momentum squared 
\[
<L^{2}>=\int d\Omega \int d^{2}p\;p^{2}\;\rho (\theta ,\phi ,p_{\theta
},p_{\phi })\; 
\]
and the angular momentum (its z-th component) 
\[
<\vec{L}>=\hat{z}\int d\Omega \int d^{2}p\;p_{\phi }\sin \theta \;\rho
(\theta ,\phi ,p_{\theta },p_{\phi })\;=<L_{z}>\hat{z} 
\]
Therefore 
\[
\rho _{1,0}(\theta ,\phi ,p_{\theta },p_{\phi })\;=\frac{3}{2\pi ^{2}}\left(
1-\frac{p_{\phi }^{2}\sin ^{2}\theta }{p^{2}}\right) K_{0}(2p)-\frac{3}{8\pi 
}\delta (p_{\theta })\delta (p_{\phi })\sin ^{2}\theta 
\]
is time independent and represents phase space density for the rigid rotor
in the state with the angular momentum squared $<L^{2}>=2$ and angular
momentum $<L_{z}>=0$, while the probability density (\ref{vjer}) is $%
P(\theta ,\phi )=\;\,\left| Y_{1,0}(\theta ,\phi )\right| ^{2}$. Similarly
the phase space density for $l=1$ and $m=1$ is 
\[
\rho _{1,1}(\theta ,\phi ,p_{\theta },p_{\phi })\;=\frac{3}{4\pi ^{2}}\left(
1+\frac{p_{\phi }^{2}\sin ^{2}\theta }{p^{2}}\right) K_{0}(2p)+\frac{%
3p_{\phi }\sin \theta }{2\pi ^{2}p}K_{1}(2p)+\frac{3}{16\pi }\left( -2+\sin
^{2}\theta \right) \delta (p_{\theta })\delta (p_{\phi }) 
\]
with the property that $P(\theta ,\phi )=\;\,\left| Y_{1,1}(\theta ,\phi
)\right| ^{2}$, $<L^{2}>=2$ and $<L_{z}>=1$.

The phase space densities for the angular momentum states $(2,m)\;;\;m=0,1,2$
were calculated as additional example. They are associated with the angular
momentum squared value $<L^{2}>=6$ and the angular momentum $<L_{z}>=0,1,2$,
respectively. They are 
\begin{eqnarray*}
\rho _{2,0}(\theta ,\phi ,p_{\theta },p_{\phi })\; &=&\frac{15}{4\pi ^{2}}%
\left( -1+\frac{p_{\phi }^{4}\sin ^{4}\theta }{p^{4}}\right) K_{0}(2p)+\frac{%
45p}{8\pi ^{2}}\left( 1-\frac{p_{\phi }^{2}\sin ^{2}\theta }{p^{2}}\right)
^{2}K_{1}(2p)+ \\
&&\frac{5}{128\pi }\left( 8-24\sin ^{2}\theta +27\sin ^{4}\theta \right)
\delta (p_{\theta })\delta (p_{\phi }) \\
\rho _{2,1}(\theta ,\phi ,p_{\theta },p_{\phi })\; &=&\frac{15}{4\pi ^{2}}%
\left( -1+2p_{\phi }\sin \theta +\frac{p_{\phi }^{2}\sin ^{2}\theta }{p^{2}}%
-2p_{\phi }^{3}\sin ^{3}\theta -\frac{2p_{\phi }^{4}\sin ^{4}\theta }{3p^{4}}%
\right) K_{0}(2p)+ \\
&&\frac{15p}{4\pi ^{2}}\left( 1-2\frac{p_{\phi }^{3}\sin ^{3}\theta }{p^{4}}-%
\frac{p_{\phi }^{4}\sin ^{4}\theta }{p^{4}}\right) K_{1}(2p)+\frac{15}{16\pi 
}\left( \sin ^{2}\theta -\frac{3}{4}\sin ^{4}\theta \right) \delta
(p_{\theta })\delta (p_{\phi }) \\
\rho _{2,2}(\theta ,\phi ,p_{\theta },p_{\phi })\; &=&\frac{15}{8\pi ^{2}}%
\left( -1+2p_{\phi }\sin \theta -2\frac{p_{\phi }^{2}\sin ^{2}\theta }{p^{2}}%
+2\frac{p_{\phi }^{3}\sin ^{3}\theta }{p^{2}}+\frac{p_{\phi }^{4}\sin
^{4}\theta }{3p^{4}}\right) K_{0}(2p)+ \\
&&\frac{15p}{16\pi ^{2}}\left( 1-4\frac{p_{\phi }\sin \theta }{p^{2}}+6\frac{%
p_{\phi }^{2}\sin ^{2}\theta }{p^{2}}+\frac{4p_{\phi }^{3}\sin ^{3}\theta }{%
3p^{4}}+\frac{p_{\phi }^{4}\sin ^{4}\theta }{p^{4}}\right) K_{1}(2p)+ \\
&&\frac{15}{32\pi }\left( 1-\sin ^{2}\theta +\frac{3}{8}\sin ^{4}\theta
\right) \delta (p_{\theta })\delta (p_{\phi })
\end{eqnarray*}

In this way phase space for the true rigid rotor was defined. It is ''true''
because only the variables that are relevant for such object were introduced.

\section{Discussion}

Analysis of angular momentum in the phase space was made, and in this
context rigid rotor was discussed. Perhaps one of the most intriguing
finding is interpretation for the partition of the kinetic energy operator,
which is in considerable disagreement with the standard one. Part of what is
considered to be radial kinetic energy is in fact contribution from the
angular momentum operator, despite the fact that it only contains radial
variable. This finding does not come as a surprise if one makes the
following observation in the traditional classical mechanics. Given
spherically symmetric probability distribution of coordinates $P(r)$ and
momenta $Q(p)$ for a free particle (say it was obtained my measuring its
position and momentum) the average kinetic energy is not zero despite the
fact that the momentum is. Likewise, the average angular momentum is zero
but the average modulus squared of it is not because 
\[
<L^{2}>=\int d^{3}r\int d^{3}p\;r^{2}(p_{\theta }^{2}+p_{\phi
}^{2})P(r)Q(p)\neq 0 
\]
Introducing the uncertainty principle does not changes this fact, except
that instead of this average having arbitrary value it has a fixed and equal
to $3/2$. Therefore a particle has always non zero modulus of angular
momentum, and its minimal value is fixed and independent of the phase space
density. It is like saying that particle always carries a minimal intrinsic
angular momentum, but in the way it is described it never manifests itself.
For all practical purpose this finding is immaterial, because standard
interpretation is self sufficient, but it becomes evident when classical
modelling is attempted.

From the phase space density for the rigid rotor one expects to obtain
momentum space probability density, which is given by 
\[
Q(\vec{p})=\int d^{3}r\;\rho (\vec{r},\vec{p}) 
\]
however one should be careful about components of the momentum variable.
Throughout the paper the components with respect to the vector $\vec{r}$
were used, because they are natural when angular momentum is analyzed. This
is because $p_{\theta }$ and $p_{\phi }$ are components along the
appropriate angular unit vectors that are perpendicular to the vector $\vec{r%
}$ and hence directly proportional to the angular momentum (they can be
called radius vector components of the momentum). However, in the momentum
space one works only with the components of the momentum vector, which for
the mentioned components this is not the case. If $p$ is modulus of the
vector $\vec{p}$ while $\theta _{p}$ and $\phi _{p}$ are its spherical
angles then 
\begin{eqnarray}
p_{r} &=&p\left( \cos \theta \;\cos \theta _{p}+\cos \left[ \phi -\phi _{p}%
\right] \sin \theta \;\sin \theta _{p}\right)  \label{pcomp} \\
p_{\theta } &=&p\left( \sin \theta \;\cos \theta _{p}-\cos \left[ \phi -\phi
_{p}\right] \cos \theta \;\sin \theta _{p}\right)  \nonumber \\
p_{\phi } &=&-p\;\sin \left[ \phi -\phi _{p}\right] \sin \theta _{p} 
\nonumber
\end{eqnarray}
which explicitly shows that the radius vector components of momentum are a
mixture of momentum and radial vector spherical coordinates. Therefore in
the phase space density they must be replaced by (\ref{pcomp}) and then the
integration over the spatial coordinates performed. Indeed for the ''soft''
rigid rotor it can be shown that one obtains for the momentum distribution
the square modulus of (\ref{mom}), but it is not clear what the outcome
would be for the true rigid rotor. For the latter it is required that $%
p_{r}=0$ and yet the expression for the phase space density would be a
function of $p$ and $\theta _{p}$ without an obvious restriction of that
kind. However, explicit expression for the momentum probability is not of
importance, it is important to be able to calculate the averages. For
example the average of the square of the Cartesian component $p_{x}$, say
with the phase space density $\rho _{2,0}$ is given by 
\[
<p_{x}^{2}>=\int d\Omega \;dp_{\theta }\;dp_{\phi }\;\left[ p_{\phi }\sin
\phi +p_{\theta }\cos \theta \cos \phi \right] ^{2}\rho _{2,0}(\theta ,\phi
,p_{\theta },p_{\phi })=\frac{22\pi }{7} 
\]
where $p_{r}=0$ was set.

\begin{figure}[tbp]
\caption{Phase space density for ''soft'' rigid rotor, for the angular
momentum indices l=m=0}
\label{fig1}
\end{figure}

%TCIMACRO{
%\TeXButton{B}{\begin{table}[tbp] \centering%
%}}%
%BeginExpansion
\begin{table}[tbp] \centering%
%
%EndExpansion
\begin{tabular}{|c|c|c|c|}
\hline
$l,$ $\mu $ & $\rho _{0,l,\mu }\cdot \pi ^{3}e^{E}$ & $<\vec{L}>$ & $<L^{2}>$
\\ \hline
0,0 & $1$ & 0 & $0+\frac{3}{2}$ \\ \hline
1,0 & $-1+2E_{z}$ & 0 & $1\ast 2+\frac{3}{2}$ \\ \hline
1,1 & $-1+E-E_{z}+2L_{z}$ & $1\;\hat{z}$ & $1\ast 2+\frac{3}{2}$ \\ \hline
2,0 & $1-\frac{2}{3}\left( E+3E_{z}\right) +\frac{1}{3}\left(
E-3E_{z}\right) ^{2}+\frac{8}{3}L^{2}-4L_{z}^{2}$ & 0 & $2\ast 3+\frac{3}{2}$
\\ \hline
2,1 & $(1-2E_{z})(1-E+E_{z}-2L_{z})$ & $1\;\hat{z}$ & $2\ast 3+\frac{3}{2}$
\\ \hline
2,2 & $-1+\frac{1}{2}\left[ E-E_{z}-2(1-L_{z})\right] ^{2}$ & $2\;\hat{z}$ & 
$2\ast 3+\frac{3}{2}$ \\ \hline
\end{tabular}
\caption{Phase space density for the ground vibrational state of harmonic oscillator for 
the first few values of angular momentum numbers. Appropriate angular
 momentum and its modulus squared are given. Definition of the variables is 
given in the text.
\label{key}} 
%TCIMACRO{
%\TeXButton{E}{\end{table}%
%}}%
%BeginExpansion
\end{table}%
%
%EndExpansion

%TCIMACRO{
%\TeXButton{B}{\begin{table}[tbp] \centering%
%}}%
%BeginExpansion
\begin{table}[tbp] \centering%
%
%EndExpansion
\begin{tabular}{|l|l|l|l|}
\hline
$l,$ $\mu $ & $\rho _{0,l,\mu }\cdot \pi ^{3}e^{E}$ & $<\vec{L}>$ & $<L^{2}>$
\\ \hline
0,0 & $1$-$\frac{1}{3}\left( 4E-2E^{2}+8L^{2}\right) $ & 0 & $0+\frac{3}{2}$
\\ \hline
1,0 & $-1+\frac{2}{5}\left( 
\begin{array}{c}
2E-E^{2}+9E_{z}-8EE_{z}+ \\ 
2E^{2}E_{z}+12L^{2}-8E_{z}L^{2}-8L_{z}^{2}
\end{array}
\right) $ & 0 & $1\ast 2+\frac{3}{2}$ \\ \hline
1,1 & $-1+\frac{1}{5}\left( 
\begin{array}{c}
13E-10E^{2}+2E^{3}-9E_{z}+8EE_{z}- \\ 
2E^{2}E_{z}+16L^{2}-8EL^{2}+8E_{z}L^{2}+10L_{z}- \\ 
8EL_{z}+4E^{2}L_{z}-16L^{2}L_{z}+8L_{z}^{2}
\end{array}
\right) $ & $1\;\hat{z}$ & $1\ast 2+\frac{3}{2}$ \\ \hline
\end{tabular}
\caption{Phase space density for the first excited vibrational state of harmonic
 oscillator for 
the first few values of angular momentum numbers. Appropriate angular
 momentum and its modulus squared are given. Definition of the variables is 
given in the text.
\label{key}} 
%TCIMACRO{
%\TeXButton{E}{\end{table}%
%}}%
%BeginExpansion
\end{table}%
%
%EndExpansion

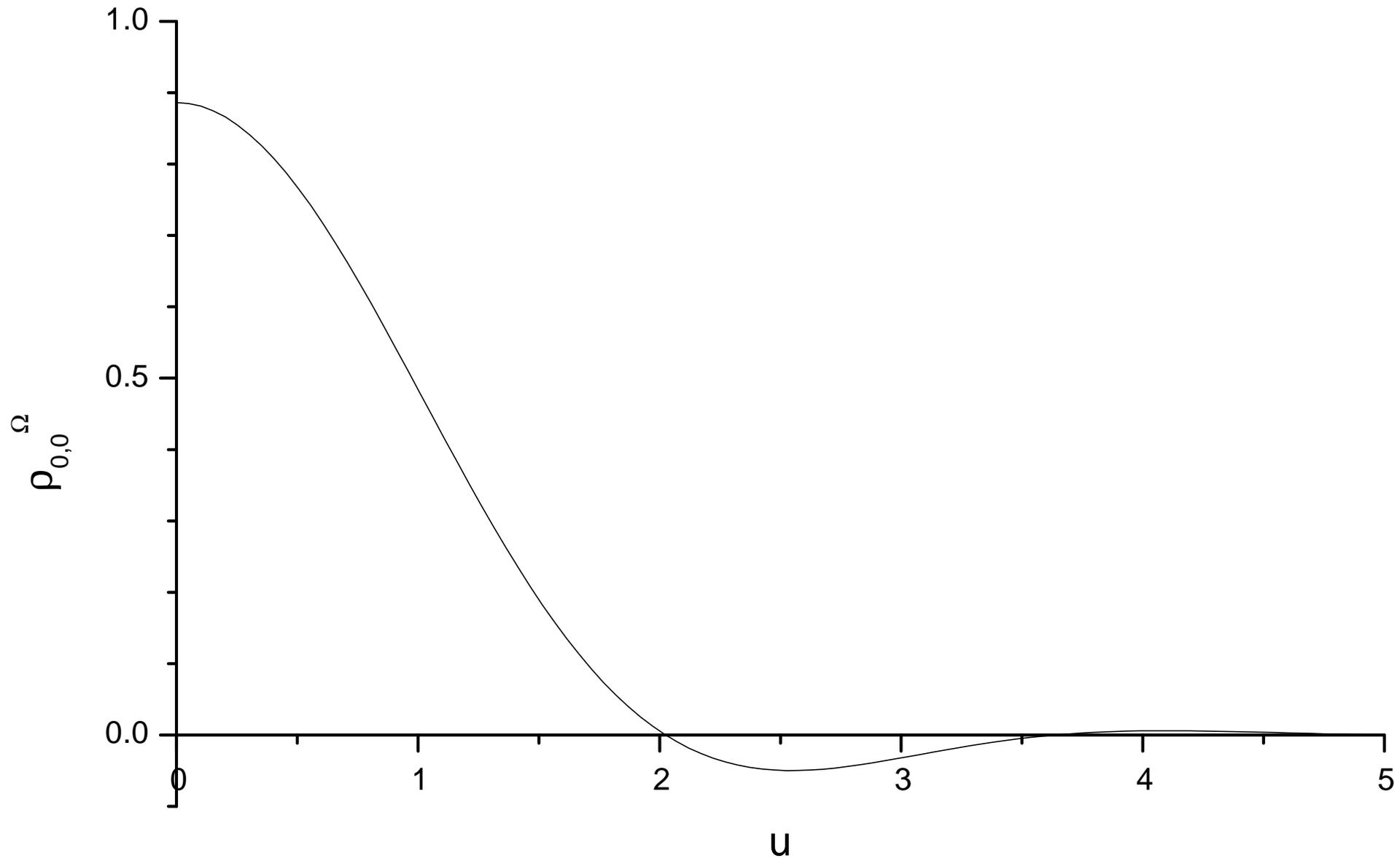
\end{document}